\documentclass[aps,eqsecnum,nofootinbib,showpacs,twocolumn]{revtex4}

\usepackage{amsmath,amsfonts,amssymb,bm}
\usepackage[dvips]{graphicx}
\usepackage{color}
\newcommand{\be}{\begin{equation}}
\newcommand{\ee}{\end{equation}}
\newcommand{\bea}{\begin{eqnarray}}
\newcommand{\eea}{\end{eqnarray}}

\newcommand{\eq}[1]{Eq.~(\ref{eq:#1})}
\newcommand{\sect}[1]{Sec.~\ref{sec:#1}}
\newcommand{\appen}[1]{App.~\ref{sec:#1}}
\newcommand{\fig}[1]{Fig.~\ref{fig:#1}}

\newcommand{\del}{\partial}
\newcommand{\eT}{\epsilon_T}

\newcommand{\vecx}{\Vec{x}}
\newcommand{\vecy}{\Vec{y}}
\newcommand{\vecq}{\Vec{q}}

\bmdefine{\bmg}{{\bm{g}}}
\bmdefine{\bmk}{{\bm{k}}}
\bmdefine{\bmx}{{\bm{x}}}
\bmdefine{\bmA}{{\bm{A}}}
\bmdefine{\bmH}{{\bm{H}}}
\bmdefine{\bmS}{{\bm{S}}}
\bmdefine{\bmphi}{{\bm{\phi}}}
\bmdefine{\bmpsi}{{\bm{\psi}}}
\bmdefine{\bmPhi}{{\bm{\Phi}}}
\bmdefine{\bmPsi}{{\bm{\Psi}}}

\newcommand{\calL}{\mathcal{L}}
\newcommand{\calO}{\mathcal{O}}

\newcommand{\calT}{\mathcal{T}}

\newcommand{\hcL}{\Hat{\calL}}
\newcommand{\Exp}[1]{{\langle\, #1\, \rangle}}
\newcommand{\mfw}{\mathfrak{w}}
\newcommand{\mfq}{\mathfrak{q}}
\newcommand{\mfA}{\mathfrak{A}}
\newcommand{\odiff}[2]{ \frac{d #1}{d #2} }
\newcommand{\odiffII}[2]{ \frac{d^2 #1}{d #2^2} }
\newcommand{\pdiff}[2]{ \frac{\partial #1}{\partial #2} }

\newcommand{\hvarphi}{\Hat{\varphi}}
\begin{document}


\title{Universality class of holographic superconductors}
\author{Kengo Maeda}
\email{maeda302@sic.shibaura-it.ac.jp}
\affiliation{Department of Engineering,
Shibaura Institute of Technology, Saitama, 330-8570, Japan}
\author{Makoto Natsuume}
\email{makoto.natsuume@kek.jp}
\affiliation{Theory Division, Institute of Particle
and Nuclear Studies, \\
KEK, High Energy Accelerator Research Organization, Tsukuba, Ibaraki,
305-0801, Japan}
\author{Takashi Okamura}
\email{tokamura@kwansei.ac.jp}
\affiliation{Department of Physics, Kwansei Gakuin University,
Sanda, Hyogo, 669-1337, Japan}
\date{\today}
\begin{abstract}
We study ``holographic superconductors" in various spacetime dimensions. We compute most of the static critical exponents in the linear perturbations and show that they take the standard mean-field values. We also consider the dynamic universality class for these models and show that they belong to model A with dynamic critical exponent $z=2$.
\end{abstract}

\maketitle

\section{Introduction}

The AdS/CFT duality \cite{Maldacena:1997re,Witten:1998qj,Witten:1998zw,Gubser:1998bc} has been a useful tool to get insights for QCD. (See Ref.~\cite{Natsuume:2007qq} for a review.) Recently, there have been many attempts to apply the AdS/CFT duality to condensed-matter physics. (See Ref.~\cite{Hartnoll:2009sz} for a review.)  
The critical phenomena may be useful to make further progress in such applications. Namely, given a field theory system, the universality class of the system may be helpful to identify the dual gravity system. In this paper, we study both the static universality class and the dynamic universality class of holographic superconductors in various dimensions.%
\footnote{
For these systems, the $U(1)$ symmetry which is spontaneously broken is a global $U(1)$ symmetry, so it may be more appropriate to use the word holographic superfluids \cite{Herzog:2008he,Basu:2008st}. For the sake of brevity, we keep using the word holographic superconductors.}

A holographic superconductor is a solution of an Einstein-Maxwell-scalar system  \cite{Gubser:2008px,HHH:2008a,HHH:2008b} or an Einstein-Yang-Mills system \cite{Gubser:2008zu}. 
There are two branches of the solution. For $T>T_c$, the solution is the standard $(p+2)$-dimensional Reissner-Nordstr\"{o}m-AdS black hole (RN-AdS$_{p+2}$) with scalar $\Psi=0$. For $T<T_c$, the solution becomes unstable and is replaced by a charged black hole with a scalar ``hair." The analytic form of the solution is unknown though, so one often uses numerical computations or uses the probe approximation, where the backreaction of the scalar field on the metric can be neglected. 

The analysis of these papers indicates that a holographic superconductor has the conventional mean-field behavior, but one had better compute all critical exponents of the system to confirm this. 
This is our main purpose. One might argue that it is not necessary to compute all critical exponents since they are related by scaling relations. However, it is not {\it a priori} obvious that a holographic superconductor obeys the scaling relations since it is not a standard statistical system but is a gravity system. Thus, computing these critical exponents is also important in order to check the AdS/CFT duality. 

In this paper, we focus on the high-temperature phase, which simplifies the analysis. We study the linear perturbations of the bulk equations of motion. Our main results are summarized as follows:
\begin{enumerate}
\item The static universality class of holographic superconductors is the one for the Gaussian fixed point. In other words, the static critical exponents take the standard mean-field values.
\item The dynamic universality class is model A with dynamic critical exponent $z=2$ satisfying the dynamic scaling relation $z=2-\eta$.
\end{enumerate}
The results are obtained by analytically studying the equations of motion without the probe approximation (\sect{general}). These results are also confirmed in numerical computations with the probe approximation (\sect{numerical}). 

In the static critical phenomena, there are six critical exponents $(\alpha,\beta,\gamma,\delta,\nu,\eta)$. (See \appen{review}.) We compute these exponents except $\beta$ and $\delta$. (The exponent $\beta$ appears only in the low-temperature phase, and $\delta$ is beyond the scope of the linear perturbations. See \sect{results_general}.) The critical exponents we found are natural since fluctuations are suppressed at large-$N$ so that the mean-field values become exact. Moreover, our results are independent of spatial dimensionality which is typical for mean-field results. 

In the dynamic case, the effect of diverging correlation length $\xi$ also appears in the relaxation time%
\footnote{One should not confuse this relaxation time with the one appeared in the second-order hydrodynamics \cite{Natsuume:2008ha}. The AdS/CFT duality for the second-order hydrodynamics was developed by Refs.~\cite{Heller:2007qt,Benincasa:2007tp,Baier:2007ix,Bhattacharyya:2008jc,Natsuume:2007ty,Natsuume:2007tz}. }
and in the transport coefficients, which is known as the critical slowing down. The divergence is parametrized by the dynamic critical exponent $z$. 

The dynamic critical phenomena in the AdS/CFT duality has been studied in Ref.~\cite{ref:DCP_our_1st} using single R-charge black holes in various dimensions \cite{Behrndt:1998jd,Kraus:1998hv,Cvetic:1999xp}. The order parameter of the transition is the charge density $\rho$. Since it is a conserved quantity, the simplest possibility for the dynamic universality class is model B according to the classification by Hohenberg and Halperin \cite{hohenberg_halperin}, and indeed it has been shown to be model B. 

For the holographic superconductors, the order parameter of the phase transition is the scalar condensate $\langle {\cal O} \rangle$ which is dual to the scalar field $\Psi$. This is not a conserved quantity, so the model should belong to model A.

We set up our notations and conventions in \sect{review} and set up the perturbation equation in \sect{perturbative_EOM}. In \sect{general}, we formally solve the perturbation equation as a double-series expansion in $(\omega, q)$, where $\omega, q$ are the frequency and the wave number, respectively. We argue that, rather generically, the values of these exponents can be understood simply in terms of a double-series expansion, and we see no sign which indicates the expansion breaks down. (If the exponents took the different values, this should appear as the break-down of the expansion somewhere.) Our argument is rather generic so that it may cover other, yet to be discovered, systems with the mean-field behavior. However, the argument in \sect{general} relies on certain assumptions, so in \sect{numerical} we numerically solve the perturbation equation in the probe approximation and obtain the exponents for the special case of the RN-AdS$_5$ solution with scalar mass $l^2 m^2 =-3$. In \appen{critical}, we review the basics of critical phenomena. In \appen{exp-crit_pt}, we show the anomalous dimension $\eta=0$ using a similar technique as in  \sect{general}.

\section{Holographic superconductors}
\label{sec:review}

In order to study the critical phenomena of holographic superconductors, we consider a $(p+2)$-dimensional Einstein-Maxwell-scalar system:
%
\begin{align}
  & \frac{\calL}{\sqrt{- g}}
  = R - 2 \Lambda
  - \frac{ F^{\mu\nu} F_{\mu\nu} }{4}
  - \left\vert D \Psi \right\vert^2 - V(| \Psi |)~,
\label{eq:action}
\end{align}
where the covariant derivative $D_\mu$, a cosmological constant $\Lambda$, and
the scalar potential $V(| \Psi |)$ are given by
\begin{align}
  & D_\mu := \nabla_\mu - i e A_\mu~,
\label{eq:def-covariant_deri} \\
  & 2 \Lambda
  = - \frac{p (p + 1)}{l^2}~,
\label{eq:def-Lambda} \\
  & V = m^2 | \Psi |^2~.
\label{eq:def-potential_V}
\end{align}
The equations of motion are given by
\begin{subequations}
\begin{align}
   R_{\mu\nu}
  &= \frac{1}{2} F_{\mu\lambda}\, F_\nu{}^\lambda
  + \big( D_{(\mu} \Psi \big)^\dagger \big( D_{\nu)} \Psi \big)
\nonumber \\
  &+ \frac{g_{\mu\nu}}{p}
    \left( 2 \Lambda - \frac{F^2}{4} + V \right)~,
\label{eq:def-gravity_eq} \\
   0
  &= \nabla_\nu F^{\mu\nu} - j^\mu
  = \frac{1}{\sqrt{-g}} \partial_\nu
    \left( \sqrt{-g} F^{\mu\nu} \right) - j^\mu~,
\label{eq:def-EM_eq} \\
   0
  &= D^\mu D_\mu \Psi - \frac{V'(| \Psi |)}{2 |\Psi|} \Psi~,
\label{eq:def-scalar_eq}
\end{align}
\end{subequations}
where the current $j$ is defined by
\begin{align}
  & j^\mu := \frac{\delta \calL_\Psi}{\delta A_\mu}
   = i e g^{\mu\nu} \Big[ \big( D_\nu \Psi \big)^\dagger \Psi
   - \Psi^\dagger \big( D_\nu \Psi \big) \Big]~.
\label{eq:def-j}
\end{align}
%

\subsection{Thermal equilibrium solution}

We study a static asymptotically AdS black hole with planar horizon, and we take the following ansatz:
\begin{subequations}
\label{eq:bg_ansatz}
\begin{align}
   ds_{p+2}^2
  &= \frac{l^2}{u^2}\, \left( - \frac{\zeta^2\, f(u)}{H^2(u)}\, dt^2
  + \zeta^2\, H^{\frac{2}{p-1}}(u)\, d\vecx_p{}^2 \right.
\nonumber \\
  &\hspace*{1.5truecm} \left.
  + \frac{H^{\frac{2}{p-1}}(u)}{f(u)}\, du^2 \right)~,
\label{eq:bg_metricI} \\
   A_\mu
  &= - \bmPhi(u)\, (dt)_\mu~,
\label{eq:bg_A} \\
   \Psi
  &= \bmPsi(u)~,
\label{eq:bg_Psi}
\end{align}
\end{subequations}
where $\zeta$ is a constant which is related to the black hole temperature [See \eq{bg-TBH}]. Without loss of generality, one can choose
\be
f(u=0) = H(u=0) = 1~.
\label{eq:bg-bc_boundary-I}
\ee
This $u=0$ is the AdS boundary. The horizon is a solution of $f(u)=0$, and we can set that the horizon is located at $u=1$ by an appropriate scaling. 

The ansatz is partly motivated by R-charged black holes \cite{Behrndt:1998jd,Kraus:1998hv,Cvetic:1999xp}. When $p=2,3$, the ansatz corresponds to the RN-AdS$_{p+2}$ limit of R-charged black holes. (The $p=2$ and $p=3$ R-charged black holes have four $U(1)$ and three $U(1)$ charges respectively, and the RN-AdS limit corresponds to the equal-charge cases.)

The equations of motion reduce to
\begin{subequations}
\label{eq:bg_eq}
\begin{align}
   0
  &= (\text{Einstein equation})~,
\label{eq:bg-grav_eq-I} \\
   0
  &= \left( \odiff{}{u} \frac{f}{u^p} \odiff{}{u}
  - \frac{ l^2 m^2 H^{ \frac{2}{p-1} } }{u^{p+2}}
  + \frac{ e^2 H^{ \frac{2 p}{p-1} } }{\zeta^2 u^p f} \bmPhi^2 \right)
    \bmPsi~,
\label{eq:bg-scalar_eq-I} \\
   0
  &= \left( \frac{u^{p-2}}{H^{2}}
    \odiff{}{u} \frac{H^{2}}{ u^{p-2} } \odiff{}{u}
  - \frac{2 l^2 e^2 H^{ \frac{2}{p-1} } }{u^2 f}
  \left\vert \bmPsi \right\vert^2 \right) \bmPhi~,
\label{eq:bg-EM_eq-t-I} \\
   0
  &=\odiff{\bmPsi^\dagger}{u} \bmPsi
  - \bmPsi^\dagger \odiff{\bmPsi}{u}~.
\label{eq:bg-EM_eq-u-I}
\end{align}
\end{subequations}
Equation~(\ref{eq:bg-EM_eq-u-I}) implies that the phase of $\bmPsi$ must be constant so that one can set $\bmPsi$ to be real without loss of generality.

The background solution is obtained from Eqs.~(\ref{eq:bg_eq}) by imposing (i) the regularity condition at the horizon and (ii) the asymptotically AdS condition. The former is given by
\begin{subequations}
\label{eq:bg-bc_horizon}
\begin{align}
  & f(u=1) = \bmPhi(u=1) = 0~,
\label{eq:bg-bc_horizon-I} \\
  & H(u=1),~\bmPsi(u=1) = \text{const.}~,
\label{eq:bg-bc_horizon-II}
\end{align}
\end{subequations}
and the latter is given by \eq{bg-bc_boundary-I} and 
\begin{subequations}
\label{eq:bg-bc_boundary}
\begin{align}
  & \bmPsi(u)
  \sim \bmpsi^{(-)} u^{\Delta_-} + \bmpsi^{(+)} u^{\Delta_+}~,
\label{eq:bg-bc_boundary-II} \\
  & \bmPhi(u) \sim \mu - C u^{p-1}~,
\label{eq:bg-bc_boundary-III}
\end{align}
\end{subequations}
where $\Delta_\pm$ are defined by
\begin{align}
  & \Delta_\pm := \frac{p + 1}{2} \pm
  \sqrt{ \left( \frac{p + 1}{2} \right)^2 + l^2 m^2 }~.
\label{eq:def-Delta}
\end{align}
Here, $\mu$ is the R-charge chemical potential and
$C$ corresponds to the R-charge density.


References~\cite{HHH:2008a,HHH:2008b} have solved Eqs.~(\ref{eq:bg_eq}) by imposing the boundary condition $\bmpsi^{(-)} = 0$ at the AdS boundary, and they found that the solution has two ``phases": 
(i) a black hole solution with no scalar hair at high temperatures and (ii) a black hole solution with a scalar hair at low temperatures. 
According to the standard AdS/CFT dictionary, $\bmpsi^{(+)}$ corresponds to the expectation value of the scalar operator dual to the scalar field, while $\bmpsi^{(-)}$ corresponds to the dual source.%
\footnote{
When $-(p+1)^2/4<l^2m^2<-(p+1)^2/4+1$, both operators are normalizable \cite{Klebanov:1999tb} so that we have two choices of the order parameter, $\langle {\cal O}_1 \rangle$ and $\langle {\cal O}_2 \rangle$ in the notations of Refs.~\cite{HHH:2008a,HHH:2008b}. In those cases, we consider only the operator $\langle {\cal O}_2 \rangle$.
}
This suggests that the hairy solution with $\bmpsi^{(+)} \neq 0$ corresponds to a spontaneous condensation of the dual operator. Furthermore, the hairy solution has the expected behavior for a superconducting phase, {\it i.e.},
(i) the divergence of the R-charge (DC) conductivity and 
(ii) an energy gap proportional to the size of the condensate.

\subsection{High-temperature phase}

At high temperatures, the black hole has no scalar hair~($\bmPsi = 0$), and the solution is given by the RN-AdS solution:
\begin{subequations}
\label{eq:RNAdS}
\begin{align}
  & H = 1 + \kappa u^{p-1}~,
\label{eq:bg-H} \\
  & f = H^{ \frac{2 p}{p-1} }
  - ( 1 + \kappa )^{ \frac{2 p}{p-1} } u^{p+1}~,
\label{eq:bg-f} \\
  & A_{t}
  = \mu \left( 1 - \frac{1 + \kappa}{H}\, u^{p-1} \right)~.
\label{eq:bg-At}
\end{align}
\end{subequations}
%
This background~(\ref{eq:bg_ansatz}) with Eqs.~(\ref{eq:RNAdS}) is parametrized by three parameters $\kappa$, $\mu$, and $\zeta$. The RN-AdS black hole may be parametrized by the temperature $T$ and the chemical potential $\mu$ (in the grand canonical ensemble). Thus, only two among three parameters are independent. In fact, they are related by%
\begin{align}
  & \mu
  = l \zeta \sqrt{ \frac{2 p}{p-1} }~
    \kappa^{1/2} (1 + \kappa)^{ \frac{1}{p-1} }~.
\label{eq:bg-mu}
\end{align}
Clearly, the Schwarzschild-AdS solution (SAdS) corresponds to $\kappa=0$. 
The black hole temperature is related to $\zeta$:
\begin{align}
   2 \pi T
  &= \zeta (1 + \kappa)^{ \frac{1}{p-1} }
  \frac{ p + 1 - (p - 1) \kappa }{2}~.
\label{eq:bg-TBH}
\end{align}
In particular, when $\kappa=0$, $\zeta$ has a simple relation with the SAdS temperature $T_0$:
\begin{align}
  & \zeta = \frac{4}{p + 1} \pi T_0~.
\label{eq:xi-T0}
\end{align}
%
The charge density $\rho$ of the RN-AdS black hole is given by
\begin{align}
  & l \rho
  = (p - 1) ( 1 + \kappa ) (l \zeta)^{p-1}\, \mu~.
\label{eq:def-rho}
\end{align}
%


\section{Perturbation equation at high $T$}
\label{sec:perturbative_EOM}

Our aim is to investigate both the static and the dynamic critical phenomena for holographic superconductors. 
The basic information about these phenomena such as critical exponents
can be read off from the two-point correlation function.
In the AdS/CFT duality, the correlation function of operator $\calO$ in the boundary theory can be obtained from the bulk perturbation of the dual scalar field.

We study the critical phenomena of holographic superconductors approaching from high temperature. This has two advantages: (i) The background solution is analytically known; (ii) The scalar perturbation $\psi := \Psi - \bmPsi$ decouples from the electromagnetic and  gravitational perturbations because the background solution has no scalar hair.

Under the ansatz
$\psi = \psi_{\mfw, \mfq}(u) e^{- i \omega t + i q x}$,
$\psi_{\mfw, \mfq}(u)$ obeys
\begin{align}
   0
  &= \left( u^p \odiff{}{u} \frac{f}{u^p} \odiff{}{u}
    + \frac{ H^{ \frac{2 p}{p-1} } }{f} \left( \mfw + \mfA \right)^2
  \right.
\nonumber \\
  &\left. \hspace{2.5truecm}
  - \mfq^2
  - l^2 m^2 \frac{ H^{ \frac{2}{p-1} } }{u^2} \right)
    \psi_{\mfw, \mfq}(u)~,
\label{eq:perturb-scalar_eq}
\end{align}
where we introduced the following dimensionless quantities:
\begin{align}
  & \mfw := \frac{\omega}{\zeta}
  = \frac{(p + 1) \omega}{4 \pi T_0}~,
& & \mfq := \frac{| q |}{\zeta}
  = \frac{(p + 1) | q |}{4 \pi T_0}~,
\label{eq:def-mfw}
\end{align}
and 
\begin{align}
  & \mfA := \frac{e A_t}{\zeta}
  = \sigma \left( \frac{1}{1 + \kappa} - \frac{u^{p-1}}{H} \right)~,
\label{eq:def-mfA} \\
  & \sigma 
  := (1 + \kappa ) \frac{e \mu}{\zeta}~.
\label{eq:def-sigma}
\end{align}
Roughly speaking, $\sigma \propto \mu/T$. Our background may be written by $T$ and $\mu$, so $\sigma$ is a unique dimensionless parameter to characterize the background. 
Also, note that $\sigma$ and $\kappa$ are related by Eq.~(\ref{eq:bg-mu}):
\begin{align}
  & \sigma
  = \sqrt{ \frac{2 p}{p-1} }~(l e) \kappa^{1/2}
    (1 + \kappa)^{\frac{p}{p-1}}~.
\label{eq:sigma-kappa}
\end{align}

Equation~(\ref{eq:perturb-scalar_eq}) is characterized by six dimensionless parameters:
\begin{itemize}
\item three parameters $p$, $(l m)$, and $(l e)$ which parametrize the action,
\item a parameter $\sigma$ (or $\kappa$) which characterizes the background, 
\item two parameters $\mfw$ and $\mfq$ which parametrize the perturbation.
\end{itemize}

The probe approximation often simplifies the analysis. In our formulation, the probe approximation is obtained by taking $e \to \infty$ keeping $\sigma$ fixed.
Equation~(\ref{eq:sigma-kappa}) tells that $\kappa \propto (l e)^{-2} \to 0$ in this limit, namely the background (\ref{eq:bg_ansatz}) becomes the SAdS solution. This limit may superficially look different from the conventional one employed in Ref.~\cite{HHH:2008a}, where one takes $e \to \infty$ keeping $(e \bmPsi)$ and $(e \bmPhi)$ fixed. But both of these limits yield the SAdS solution. 

In this approximation, Eq.~(\ref{eq:perturb-scalar_eq}) is characterized by five parameters: $p, (l m), \sigma$ (with $\kappa=0$), $\mfw$, and $\mfq$. We will employ the probe approximation for numerical analysis in \sect{numerical}, but we will not employ the approximation for the analytic argument in \sect{general}.

Near the horizon $u \sim 1$, Eq.~(\ref{eq:perturb-scalar_eq})  becomes 
\begin{align}
   0
  &= \left( f \pdiff{}{u} f \pdiff{}{u}
  + (1 + \kappa)^{ \frac{2 p}{p-1} } \mfw^2 \right)
  \psi_{\mfw, \mfq}(u)~,
\label{eq:perturb-scalar_eq-near_horizon}
\end{align}
so the solution is given by 
$\psi_{\mfw, \mfq} \sim (1 - u)^{ \pm i \frac{\omega}{4 \pi T} }$.
We impose the ``incoming wave" boundary condition at the horizon, which corresponds to the retarded condition:
\begin{align}
  \psi_{\mfw, \mfq}
  \sim (1 - u)^{ - i \frac{\omega}{4 \pi T} }~.
\label{eq:behavior-near_horizon}
\end{align}

Near the AdS boundary $u \sim 0$,
Eq.~(\ref{eq:perturb-scalar_eq}) becomes
\begin{align}
   0
  &= \left( u^p \pdiff{}{u} u^{-p} \pdiff{}{u}
  - \frac{l^2 m^2}{u^2} \right) \psi_{\mfw, \mfq}(u)~,
\label{eq:perturb-scalar_eq-near_boundary}
\end{align}
and $\psi_{\mfw, \mfq}$ has the same fall-off behavior as the background $\bmPsi$:
\begin{align}
  & \psi_{\mfw, \mfq}(u)
  \sim \psi_{\mfw, \mfq}^{(-)} u^{\Delta_-}
  + \psi_{\mfw, \mfq}^{(+)} u^{\Delta_+}~.
\label{eq:behavior-near_boundary}
\end{align}

The critical phenomena near the second-order phase transition can be extracted from the two-point correlation function or the response function. In the AdS/CFT duality, the source term corresponds to $\psi_{\mfw, \mfq}^{(-)}$, and the dual order parameter expectation value $\Exp{\calO_{\mfw, \mfq}}$ corresponds to $\psi_{\mfw, \mfq}^{(+)}$, so the response function is given by
\begin{align}
   \chi_{\mfw, \mfq}
  &:= \left. \frac{ \delta \Exp{ \calO_{\mfw, \mfq} } }
                  { \delta \psi^{(-)}_{\mfw, \mfq} }
      \right\vert_{ \psi^{(-)}=0 }
  \propto \frac{ \psi^{(+)}_{\mfw, \mfq} }{ \psi^{(-)}_{\mfw, \mfq} }~.
\label{eq:def-response_func}
\end{align}
Thus, our task is to solve Eq.~(\ref{eq:perturb-scalar_eq}) under
the boundary condition (\ref{eq:behavior-near_horizon}), 
obtain the coefficients $\psi_{\mfw, \mfq}^{(\pm)}$ in \eq{behavior-near_boundary},
and study the behavior of the response function $\chi_{\mfw, \mfq} \propto
\psi_{\mfw, \mfq}^{(+)}/\psi_{\mfw, \mfq}^{(-)}$.

\section{General structure of response function by $(\mfw, \mfq)$-expansion}
\label{sec:general}

First, we formally solve the perturbation equation (\ref{eq:perturb-scalar_eq}) as a double-series expansion in $(\mfw, \mfq)$. We demonstrate that the critical exponents for holographic superconductors can be understood simply in terms of a double-series expansion. 

In order to implement the ``incoming wave" boundary condition (\ref{eq:behavior-near_horizon}), it is convenient to introduce a new variable $\varphi$ as
\begin{align}
  & \psi_{\mfw, \mfq}(u)
  =: \calT(u) \varphi_{\mfw, \mfq}(u)~.
\label{eq:def-varphi}
\end{align}
The choice of the function $\calT(u)$ is arbitrary as long as it satisfies 
\begin{align}
  & \calT(u)
  \sim ~(1 - u)^{ - i \frac{\omega}{4 \pi T} }
& & \text{~~for~~} u \sim 1~,
\label{eq:def-calT}
\end{align}
but we choose
\begin{align}
  & \calT
  = \exp\left[ i \int^u_0 du \frac{ H^{\frac{p}{p-1}} }{f}
    (\mfw + \mfA) \right]~.
\label{eq:def-calT-exmpl}
\end{align}
Then, the ``incoming wave" boundary condition at the horizon corresponds to $\varphi_{\mfw, \mfq}(u=1) = \text{const.}$, and Eq.~(\ref{eq:perturb-scalar_eq}) becomes
\begin{subequations}
\label{eq:perturb_eq-general}
\begin{align}
  & 0
  = \left( \odiffII{}{u} + B_1(u) \odiff{}{u} + B_0(u) \right)
    \varphi_{\mfw, \mfq}(u)~,
\label{eq:perturb_eq-general-varphi} \\
  & B_1(u)
  = \odiff{}{u} \ln \frac{f}{u^p}
  + 2 i\, \frac{ H^{\frac{p}{p-1}} }{f}\, (\mfw + \mfA)~,
\label{eq:B_1-exmpl} \\
  & B_0(u)
  = - \frac{1}{f} \left( \mfq^2
    + l^2 m^2\, \frac{ H^{ \frac{2}{p-1} } }{u^2} \right)
\nonumber \\
  &\hspace*{2.0truecm}
  + i\, \frac{u^p}{f} \odiff{}{u}
      \left( \frac{ H^{\frac{p}{p-1}} }{u^p}\, (\mfw + \mfA) \right)~.
\label{eq:B_0-exmpl}
\end{align}
\end{subequations}
Near the AdS boundary, $\varphi_{\mfw, \mfq}$ behaves as
\begin{align}
  & \varphi_{\mfw, \mfq}(u)
  \sim \varphi_{\mfw, \mfq}^{(-)}~u^{\Delta_-}
  + \varphi_{\mfw, \mfq}^{(+)}~u^{\Delta_+}~,
\label{eq:near_boundary-general}
\end{align}
where $\Delta_\pm~(\Delta_- < \Delta_+)$ is defined in Eq.~(\ref{eq:def-Delta}).

Let us solve Eq.~(\ref{eq:perturb_eq-general-varphi}) as a double-series expansion in $(\mfw, \mfq)$:
\begin{align}
   \varphi_{\mfw, \mfq}(u)
  &= \varphi_{0}(u) + \mfw\, \varphi_{(1, 0)}(u)
  + \mfq^2 \varphi_{(0, 1)}(u)
\nonumber \\
  &+ O(\mfw^2, \mfw \mfq^2, \mfq^4)~.
\label{eq:varphi-exp-general}
\end{align}
Each variable obeys the following differential equations:
\begin{subequations}
\begin{align}
  & \hcL \varphi_{0}(u)
  := \left[ \odiffII{}{u} + C_1(u) \odiff{}{u} + C_0(u) \right]
  \varphi_{0}(u) = 0~,
\label{eq:perturb_eq-general-exp_zeroth} \\
  & \hcL \varphi_{(1, 0)}(u) = J_{(1, 0)}(u)~,
\label{eq:perturb_eq-general-exp_omega} \\
  & \hcL \varphi_{(0, 1)}(u) = J_{(0, 1)}(u)~,
\label{eq:perturb_eq-general-exp_k2}
\end{align}
\end{subequations}
where $C_1(u)$, $C_0(u)$, $J_{(1,0)}(u)$, and $J_{(0,1)}(u)$
are defined by
\begin{subequations}
\label{eq:C-exmpl}
\begin{align}
  & C_1(u)
  = \odiff{}{u} \ln \frac{f}{u^p}
  + 2 i \frac{ H^{\frac{p}{p-1}} }{f} \mfA~,
\label{eq:C_1-exmpl} \\
  & C_0(u)
  = - l^2 m^2 \frac{ H^{ \frac{2}{p-1} } }{u^2 f}
  + i\, \frac{u^p}{f} \odiff{}{u}
      \left( \frac{ H^{\frac{p}{p-1}} }{u^p}\, \mfA \right)~,
\label{eq:C_0-exmpl}
\end{align}
\end{subequations}
and
\begin{subequations}
\label{eq:J-exmpl}
\begin{align}
  & J_{(1,0)}(u)
  = - i \frac{ H^{\frac{p}{p-1}} }{f} \left[ 2 \odiff{}{u}
  + \odiff{}{u} \left(
    \ln \frac{ H^{\frac{p}{p-1}} }{u^p} \right)
    \right] \varphi_0~,
\label{eq:J_1_0-exmpl} \\
  & J_{(0,1)}(u)
  = \frac{\varphi_0}{f}~.
\label{eq:J_0_1-exmpl}
\end{align}
\end{subequations}
%

\subsection{Zeroth order in $(\mfw, \mfq)$-expansion}

From the boundary condition, the zeroth-order solution $\varphi_0(u)$ must be regular at the horizon, and it must satisfy
\begin{align}
  & \left. \frac{\varphi_{0}'}{\varphi_{0}} \right\vert_{u=1}
  = - \left. \frac{C_0}{C_1} \right\vert_{u=1}
  =  \left.
  \frac{ l^2 m^2 H^{\frac{2}{p-1}} - i H^{ \frac{p}{p-1} } \mfA' }
       {f'} \right\vert_{u=1}~.
\label{eq:behavior-near_horizon-zeroth}
\end{align}
This determines the boundary condition for $\varphi_0$ up to an overall constant, so we shall set $\varphi_{0}(u=1) = 1$.

Suppose that one solves the differential equation (\ref{eq:perturb_eq-general-exp_zeroth}) under this boundary condition by integrating out from the horizon to the AdS boundary. Then, one can determine the behavior of $\varphi_0(u)$ as 
\begin{align}
  & \varphi_{0}(u)
  \sim \varphi_{0}^{(-)}~u^{\Delta_-}
  + \varphi_{0}^{(+)}~u^{\Delta_+}~,
\label{eq:near_boundary-zeroth}
\end{align}
and determine the ratio $\varphi_{0}^{(+)}/\varphi_{0}^{(-)}$. This ratio is proportional to the thermodynamic susceptibility of the order parameter, $\chi$:
\begin{align}
  & \chi \propto \psi_{0}^{(+)}/\psi_{0}^{(-)}
  = \varphi_{0}^{(+)}/\varphi_{0}^{(-)}~
\label{eq:def-chi}
\end{align}
because the thermodynamic susceptibility $\chi$ is the response function (\ref{eq:def-response_func}) for the stationary homogeneous source $\chi = \chi_{ \mfw=\mfq=0 }$. 

In the high-temperature phase $T>T_c$, it is possible to show that the order parameter $\varphi_{0}^{(+)}$ vanishes if there is no source $\varphi_{0}^{(-)}$. Thus, $\varphi_{0}^{(+)}$ is proportional to $\varphi_{0}^{(-)}$ and $\chi$ is finite,  which is the behavior similar to paramagnets.

On the other hand, at the critical point, the spontaneous condensation occurs, so one must have $\varphi_{0}^{(-)} |_{T_c} = 0$ and $\varphi_{0}^{(+)} |_{T_c} \ne 0$; we assume that this is the case. This is nothing but the divergence of the thermodynamic susceptibility $\chi$ at the critical point. Then, the critical exponent of the thermodynamic susceptibility can be extracted from the behavior of $\varphi_{0}^{(-)}$. As $T$ approaches $T_c$ from the above, we assume
\begin{align}
  & \lim_{T \searrow T_c} \psi_{0}^{(-)} = 0~,
& & \lim_{T \searrow T_c} \psi_{0}^{(+)} = \text{finite}~.
\label{eq:aympto_conjecture}
\end{align}
So, denote the deviation from the critical temperature $T_c$ by $\epsilon_T := T/T_c - 1 > 0$, and write $\psi_{0}^{(-)}$ as
\begin{align}
  & \psi_{0}^{(-)} = \varphi_{0}^{(-)} \propto \epsilon_T^\gamma
& & ( \gamma > 0 )~.
\label{eq:psi_zero-eps_behavior-general}
\end{align}
Then, the thermodynamic susceptibility behaves as
\begin{align}
  & \chi \propto \varphi_{0}^{(+)}/\varphi_{0}^{(-)}
  \propto \epsilon_T^{-\gamma}~,
\label{eq:exponent-chi}
\end{align}
namely $\gamma$ in Eq.~(\ref{eq:psi_zero-eps_behavior-general}) is indeed the critical exponent of the thermodynamic susceptibility.

\subsection{First order in $(\mfw, \mfq)$-expansion}

It is easy to solve Eqs.~(\ref{eq:perturb_eq-general-exp_omega}) and 
(\ref{eq:perturb_eq-general-exp_k2}) formally.
Let $\hvarphi_0$ be a zeroth-order solution which is linearly independent from $\varphi_{0}$. Then, one can construct the Green function of the operator $\hcL$
which satisfies the regularity condition at the horizon:
\begin{align}
   G(u, u'\,)
  &:= \theta( u - u'\,) \varphi_0(u) \hvarphi_0(u'\,)
\nonumber \\
  &+ \theta( u' - u ) \hvarphi_0(u) \varphi_0(u'\,)~,
\label{eq:Green_func-general} \\
   W(u)
  &:= \odiff{\varphi_0(u)}{u} \hvarphi_0(u)
  - \varphi_0(u) \odiff{\hvarphi_0(u)}{u}~,
\label{eq:def-W-general}
\end{align}
where the Green function $G$ and the Wronskian $W$ satisfy
\begin{align}
  & \hcL_{u} G(u, u'\,) = W(u) \delta( u - u'\,)~,
\label{eq:Green_eq-general} \\
  & \odiff{}{u} W(u) + C_1(u) W(u) = 0~.
\label{eq:W_eq-general}
\end{align}
Using the Green function (\ref{eq:Green_eq-general}), the solution of \eq{perturb_eq-general-exp_omega} is expressed formally as
\begin{align}
   \varphi_{(1,0)}(u)
  &= \int^1_0 du' G(u, u'\,) \frac{J_{(1,0)}(u'\,)}{W(u'\,)}~,
\nonumber \\
  &= \varphi_0(u) \int^u_0 du'
    \hvarphi_0(u'\,) \frac{J_{(1,0)}(u'\,)}{W(u'\,)}
\nonumber \\
  &+ \hvarphi_0(u) \int^1_u du'
    \varphi_0(u'\,) \frac{J_{(1,0)}(u'\,)}{W(u'\,)}~,
\label{eq:solution-general-exp}
\end{align}
and similarly for \eq{perturb_eq-general-exp_k2}.

We also assume that the first-order solutions have the same fall-off behavior as Eq.~(\ref{eq:behavior-near_boundary}). Then, 
\begin{align}
   \varphi_{\mfw, \mfq}(u)
  &\sim \varphi_{0}^{(-)} u^{\Delta_-} + \varphi_{0}^{(+)} u^{\Delta_+}
\nonumber \\
  &+ \mfw \left( \varphi_{(1,0)}^{(-)} u^{\Delta_-}
    + \varphi_{(1,0)}^{(+)} u^{\Delta_+} \right)
\nonumber \\
  &+ \mfq^2 \left( \varphi_{(0,1)}^{(-)} u^{\Delta_-}
    + \varphi_{(0,1)}^{(+)} u^{\Delta_+} \right)~.
\label{eq:psi-exp-general-boundary}
\end{align}
Comparing this with Eq.~(\ref{eq:near_boundary-general}), we obtain
\begin{subequations}
\label{eq:varphi-general}
\begin{align}
  & \psi_{\mfw, \mfq}^{(-)}
  = \varphi_{\mfw, \mfq}^{(-)}
  = \varphi_{0}^{(-)} + \mfw \varphi_{(1,0)}^{(-)}
  + \mfq^2 \varphi_{(0,1)}^{(-)}~,
\label{eq:varphi_m-general} \\
  & \psi_{\mfw, \mfq}^{(+)}
  = \varphi_{\mfw, \mfq}^{(+)}
  = \varphi_{0}^{(+)} + \mfw \varphi_{(1,0)}^{(+)}
  + \mfq^2 \varphi_{(0,1)}^{(+)}~.
\label{eq:varphi_p-general}
\end{align}
\end{subequations}
%

\subsection{The order parameter response function}\label{sec:results_general}

For readers' convenience, let us list our assumptions used so far: 
\begin{enumerate}
%
\item The first-order solutions $\varphi_{(1,0)}$ and $\varphi_{(0,1)}$ indeed exist,
\item They have the same fall-off behavior as Eq.~(\ref{eq:behavior-near_boundary}),
\item The singular behaviors at the critical point come only from  $\varphi_{0}^{(-)}$, {\it i.e.},  $\varphi_{0}^{(-)} |_{T_c} = 0$. In particular, we shall use 
$\varphi_{0}^{(+)} |_{T_c}, \varphi_{(1,0)}^{(-)} |_{T_c}, \varphi_{(0,1)}^{(-)} |_{T_c} \ne 0$. 
\end{enumerate}
Using the above assumptions and the results in previous subsections, the response function (\ref{eq:def-response_func}) is given by
\begin{align}
   \chi_{\mfw, \mfq}
  & \propto
    \frac{ \varphi_{0}^{(+)} + \mfw \varphi_{(1,0)}^{(+)}
          + \mfq^2 \varphi_{(0,1)}^{(+)} }
         { \varphi_{0}^{(-)} + \mfw \varphi_{(1,0)}^{(-)}
          + \mfq^2 \varphi_{(0,1)}^{(-)} }
\nonumber \\
  &\sim \frac{ \varphi_{0}^{(+)} }{ \varphi_{(0,1)}^{(-)} }~
    \frac{1}{- i c \mfw + \mfq^2 + 1/\xi^2 }~.
\label{eq:response_func-general}
\end{align}
where $c$ and $\xi$ are defined by
\begin{align}
   c
  &:= i \varphi_{(1,0)}^{(-)}/\varphi_{(0,1)}^{(-)}~,
& & \xi^2
  := \varphi_{(0,1)}^{(-)}/\varphi_{0}^{(-)}~.
\label{eq:def-xi-general}
\end{align}
In particular, the static and dynamic response function are given by
\begin{align}
  & \chi_{\mfw=0, \mfq}
  \sim \frac{ \varphi_{0}^{(+)} }{ \varphi_{(0,1)}^{(-)} }~
    \frac{1}{ \mfq^2 + 1/\xi^2 }~,
\label{eq:static_response_func-general} \\
  & \chi_{\mfw, \mfq=0}
  \sim \frac{ \varphi_{0}^{(+)} }{ \varphi_{(0,1)}^{(-)} }~
    \frac{1}{- i c \mfw + 1/\xi^2 }~.
\label{eq:dynamic_response_func-general}
\end{align}
From the response functions (\ref{eq:response_func-general}) and (\ref{eq:static_response_func-general}), we can extract the following information:
\begin{itemize}
\item 
The form of the static response function (\ref{eq:static_response_func-general}) tells that $\xi$ is indeed the correlation length, and $\xi$ diverges as $\xi \propto \epsilon_T^{-\gamma/2}$  from Eqs.~(\ref{eq:psi_zero-eps_behavior-general}) and (\ref{eq:def-xi-general}). This determines the critical exponent $\nu$ (defined by $\xi \propto \epsilon_T^{-\nu}$) as $\nu = \gamma/2$.

At the critical point ($\epsilon_T=0$), the static response function behaves
as $\chi_{\mfw=0, \mfq} \vert_{T_c} \propto \mfq^{-2}$. 
This means that the anomalous dimension $\eta$,
which is defined by $\chi_{\mfw=0, \mfq} \vert_{T_c}
\propto \mfq^{\eta-2}$, vanishes.
These critical exponents $\gamma$, $\nu$, and $\eta$ satisfy the static scaling relation
$\gamma = \nu ( 2 - \eta )$, as expected.
\vspace{0.2truecm}
\item The dynamic response function
(\ref{eq:dynamic_response_func-general}) indicates
that the homogeneous perturbation of the dual order parameter
decays on the time scale
$\tau_{\mfq=0} \sim c\, \xi^2$.
This is the critical slowing down, in which the relaxation time diverges as a power of the correlation length (if we assume $c \neq 0$ near the critical point).
The dynamic critical exponent $z$, which is defined by $\tau_{\mfq=0} \propto \xi^z$, is equal to 2. This exponent $z$ also appears in the dynamic response function at the critical point: it appears in the $\mfq$-dependence as $\tau_{\mfq} \vert_{T_c} \propto \mfq^{-z}$.
In fact, Eq.~(\ref{eq:response_func-general}) gives  $\tau_{\mfq} \sim c\, \mfq^{-2}$
at the critical point (when $\xi = \infty$).
\end{itemize}
This is our main result.

The explicit value of the exponent $\gamma$ is determined either from the numerical computation in \sect{numerical} or from an analytic argument in \appen{exp-crit_pt}: we find $\gamma = 1$. As a consequence, the static critical exponents are
\begin{align}
  & \gamma = 1~,
& & \nu = \gamma/2 = 1/2~,
& & \eta = 0~,
\label{eq:exponents}
\end{align}
which satisfy the static scaling relation $\gamma = \nu ( 2 - \eta )$. Note that our results are independent of spatial dimensionality which is typical for mean-field results. In Refs.~\cite{ref:MaedaOkamura_08,HR:2008}, the exponent $\nu$ has been computed in the low-temperature phase. We find that the exponent takes the same value in the high-temperature phase, which is typical in critical phenomena.

It is important to check the other static scaling relations
\begin{align}
  & \alpha + 2 \beta + \gamma = 2~,
& & \gamma = \beta ( \delta - 1 )~.
\label{eq:else_scaling_rel}
\end{align}
Here, $\alpha$ is the exponent for the heat capacity, $\delta$ is the exponent which gives the behavior of the order parameter at the critical point:
$\Exp{\calO_{\mfw=\mfq=0}} \propto |\psi_0^{(-)} |^{1/\delta}$.
The exponent $\beta$ determines the behavior of the order parameter in the ordered phase: 
$\Exp{\calO_{\mfw=\mfq=0}} \propto |\epsilon_T |^{\beta}$. We focus on the high-temperature phase, so we cannot obtain $\beta$, but $\beta=1/2$ from Refs.~\cite{HHH:2008a,Gubser:2008zu,HHH:2008b}. 

As we approach the critical point from the high-temperature phase, the bulk spacetime is the RN-AdS, so the heat capacity is given by the one for the RN-AdS. Thus, the heat capacity converges to a constant value as we approach the critical point, which implies $\alpha=0$. Hence, the first scaling relation in \eq{else_scaling_rel} is confirmed. 

In order to check the second scaling relation in \eq{else_scaling_rel}, one needs to compute $\delta$, but this analysis itself does not give $\delta$. The would-be value for $\delta$ is $\delta=3$, and this means that $ \psi_0^{(-)} \propto (\psi_0^{(+)})^3 $, so the linear perturbation is not enough to derive the result. (See however Ref.~\cite{Herzog:2008he}, where the authors show that the standard Landau potential fits their numerical results. This indicates $\delta=3$.)

We find that the static critical exponents take the standard mean-field values (except $\delta$). One might suspect that our argument is simply a consequence of double-series expansion. But the whole point is that the universality class of holographic superconductors can be understood in this way. If the exponents took the different values, this should appear as the breakdown of the expansion somewhere, but we see no sign of the breakdown. However, our argument relies on certain assumptions, so we numerically obtain the exponents in the next section. 

Now, let us turn to the dynamic critical exponent $z$.
We consider the linear perturbations of the bulk field.
In the dual field theory, this corresponds to the following approximations:
\begin{itemize}
\item[(a)] One ignores the couplings between the fluctuations
of the order parameter with different wave number 
(the Gaussian approximation for the fluctuations).
\item[(b)] One ignores the nonlinear couplings between
hydrodynamic modes and the fluctuations of the order parameter 
(no mode coupling).
\end{itemize}
This approximation is known as van Hove theory
(``conventional theory") of the dynamic critical phenomena,
which predicts that the dynamic critical exponent $z$
is related to the static critical exponent $\eta$ (anomalous dimension) by $z = 2 + a - \eta$.
Here, $a=0$ when the order parameter is not a conserved quantity (model~A)
and $a=2$ when it is a conserved one (model~B).

In holographic superconductors, the order parameter is the scalar condensate $\langle \calO \rangle$ which is not a conserved quantity, so we expect $z = 2 - \eta$ from van Hove theory. This is consistent with our results $\eta = 0$, $z = 2$ obtained
from the $(\mfw, \mfq)$-expansion, implying that the AdS/CFT duality also holds in the dynamic case. In other words, these black holes do obey the theory of dynamic critical phenomena.

It is interesting to see
whether $z$ obtained from nonlinear perturbations is consistent
with the properties predicted by the dynamic renormalization group analysis. 
Under nonlinear perturbations,
one generally expects (a') the violation of the Gaussian approximation and 
(b') the appearance of mode-mode coupling.
Since the dual field theory is at large-$N$, the effect of renormalization from short-wavelength fluctuations would be suppressed and we could ignore (a').
On the other hand, (b') has a substantial effect, which would change the dynamic universality class from model A (maybe to model~F which is the universality class of $^4$He). It would be interesting if one could obtain $z$ in the gravity side which is consistent with the dynamic renormalization group analysis. This could become a strong support of the AdS/CFT duality in the dynamic regime.%
\footnote{
This nonlinear effect (b') may be subleading in the $1/N$-expansion. We thank Misha Stephanov for pointing this out to us.
}

\section{Numerical results}
\label{sec:numerical}

The results in \sect{general} and \appen{exp-crit_pt} are obtained by assuming that 
(i) Eq.~(\ref{eq:perturb_eq-general}) can be solved as a double-series expansion in $(\mfw, \mfq)$, 
(ii) each other solution preserves the fall-off behavior~(\ref{eq:behavior-near_boundary}), 
and (iii) singular behaviors come only from $\varphi_{0}^{(-)}$. 
In this section, we solve Eq.~(\ref{eq:perturb_eq-general}) numerically for a particular set of parameters and show that our results based on the above assumptions are indeed correct. 

We numerically solve Eq.~(\ref{eq:perturb_eq-general}) under the boundary condition
\begin{align}
  & \left. \varphi_{\mfw, \mfq} \right\vert_{u=1} = 1~,
& & \left. \frac{\varphi_{\mfw,\mfq}'}{\varphi_{\mfw,\mfq}}
  \right\vert_{u=1}
  = - \left. \frac{B_0}{B_1} \right\vert_{u=1}~,
\label{eq:behavior-near_horizon-general}
\end{align}
and obtain the response function
$\chi_{\mfw, \mfq} \propto
\varphi_{\mfw, \mfq}^{(+)}/\varphi_{\mfw, \mfq}^{(-)}$
from the asymptotic behavior (\ref{eq:near_boundary-general}).

To simplify the analysis, we perform the numerical calculation in the probe approximation. The perturbed solution then depends on five parameters $p$, $(l m)$, $\sigma$, $\mfw$, and $\mfq$ (\sect{perturbative_EOM}). As an explicit example, we consider five-dimensional bulk spacetime ($p=3$) with the scalar mass $l^2 m^2 = -3$, which was discussed in Ref.~\cite{HR:2008}. In the probe approximation, the background~(\ref{eq:bg_ansatz}) becomes the SAdS solution, so $f = 1 - u^4$, $H = 1$.

\subsection{Critical point and thermodynamic susceptibility}

In our formulation, the dimensionless parameter $\sigma$ determines the phase structure, and let us first find the critical point $\sigma_c$. We solve 
\begin{itemize}
\item EOM: Eq.~(\ref{eq:perturb_eq-general-varphi}) with $\mfw = \mfq = 0$, or Eq.~(\ref{eq:perturb_eq-general-exp_zeroth}), 
\item Boundary conditions: \eq{behavior-near_horizon-general} at the horizon and $\varphi_0^{(-)} = 0$ (corresponding to a spontaneous condensate) at the AdS boundary.
\end{itemize}
Then, $\sigma_c$ is obtained as an eigenvalue problem under two boundary conditions. We use a numerical method developed by Horowitz and Hubeny to solve such an eigenvalue problem \cite{Horowitz:1999jd}.
The parameter is numerically found as
\footnote{
Reference~\cite{HR:2008} found that $T_c = 0.198 \times \rho_{\text{HR}}^{1/3}$ 
in units $l = e = 1$, which coincides with our numerical result.
Note that the R-charge density $\rho_{\text{HR}}$ in Ref.~\cite{HR:2008}
is related to our $\rho$ by $\rho_{\text{HR}} = \rho/2$.
}
\be
\sigma_c \sim 4.15686 
\quad \mbox{or} \quad 
T_c \sim 0.19797 \times \left( \frac{e \rho}{2l} \right)^{1/3}~.
\ee

\begin{figure}
  \includegraphics[width=8.0truecm]{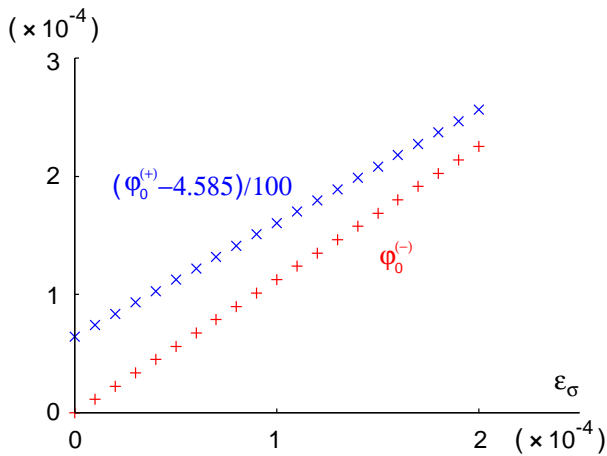}
\caption{\label{fig:behavior_psi} (color online).
  The value of $\varphi_0^{(-)}$ (``$+$", (red))
  and $(\varphi_0^{(+)} - 4.585)/100$ (``$\times$", (blue))
  as a function of $\sigma$.
  As $\sigma$ approaches the critical value $\sigma_c \sim 4.15686$,
  $\varphi_0^{(-)}$ goes to zero
  and $\varphi_0^{(+)}$ goes to a finite constant.
}
\end{figure}

\begin{figure}
  \includegraphics[width=8.0truecm]{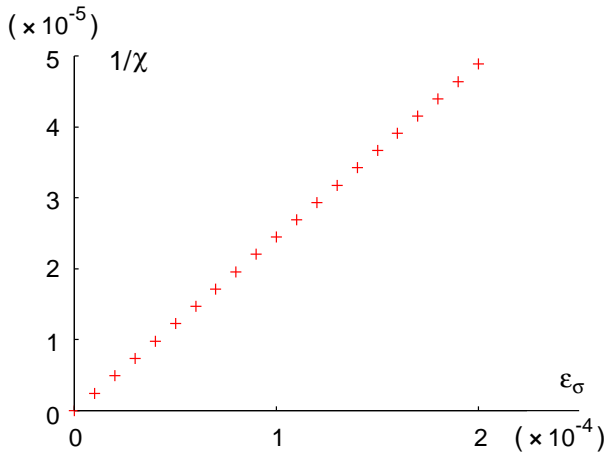}
\caption{\label{fig:thermo_chi} (color online).
  The thermodynamic susceptibility $\chi$
  as a function of $\epsilon_\sigma := 1 - \sigma/\sigma_c$.
  Plotted are $1/\chi$ for $\epsilon_\sigma = 10^{-5} n$
  ($n = 0, 1, \cdots, 20$).
}
\end{figure}

To examine the critical behavior of the thermodynamic susceptibility
$\chi$, deviate $\sigma$ from the critical point. We denote the deviation from the critical point by $\epsilon_\sigma$, where 
\begin{subequations}
\label{eq:eps_sigma}
\begin{align}
   \epsilon_\sigma
  &:= 1 - \sigma/\sigma_c \propto \epsilon_T
\label{eq:def-eps_sigma} \\
  &= 10^{-5}\, n
& & (n=0, 1, \cdots, 20)~,
\label{eq:eps_sigma-range}
\end{align}
\end{subequations}
and see how physical quantities behave under the deformation. Figure~\ref{fig:behavior_psi} shows $\varphi_0^{(\pm)}$ as a function of $\sigma$ around the critical point~(in the high-temperature phase). As expected, the critical behavior coincides
with Eq.~(\ref{eq:aympto_conjecture}). This guarantees that $\chi$ diverges near the critical point, and \fig{thermo_chi} shows that $\chi$ diverges as $\chi \propto 1/\epsilon_\sigma$, suggesting $\gamma = 1$.
In fact, the result may be fitted by a polynomial as
\begin{align}
   1/\chi
  &\sim - 1.72 \times 10^{-8} + 0.25 \times \epsilon_\sigma
  - 4.58 \times \epsilon_\sigma^2
\nonumber \\
  &\propto \epsilon_\sigma~,
\label{eq:chi_fitting}
\end{align}
so $\gamma = 1$ within numerical errors.

\subsection{Correlation length and static susceptibility}

We then check $\eta = 0$, $\nu = 1/2$. These are exponents for the static susceptibility $\chi_{\mfw=0, \mfq}$, so it is enough to consider the $\mfw = 0$ perturbation.

In order to obtain $\nu$, we solve 
\begin{itemize}
\item EOM: Eq.~(\ref{eq:perturb_eq-general-varphi}) with $\mfw = 0$, 
\item Boundary conditions: \eq{behavior-near_horizon-general} at the horizon and $\varphi_{\mfw=0,\mfq}^{(-)} = 0$ (corresponding to no deformation) at the AdS boundary.
\end{itemize}
This procedure gives the eigenvalue $\mfq_*$, which corresponds to the pole of the static susceptibility. The pole gives the correlation length $\xi$ by $\xi^2:=- 1/\mfq_*^2$.
\begin{figure}
  \includegraphics[width=8.0truecm]{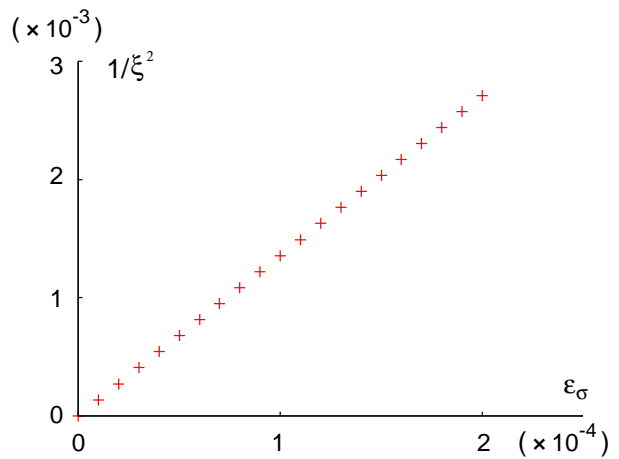}
\caption{\label{fig:xi_sigma} (color online).
  The correlation length $\xi^2$ defined by $\xi^2 := - 1/\mfq_*^2$.
  Plotted are $1/\xi^2$
  for $\epsilon_\sigma = 10^{-5} n$
  ($n = 0, 1, \cdots, 20$).
}
\end{figure}%
Figure~\ref{fig:xi_sigma} shows $\xi^2$ with the interval 
$\Delta\epsilon_\sigma = 10^{-5}$
toward the critical value $\sigma_c$, which suggests $\xi \propto \epsilon_\sigma^{-1/2}$. The result may be fitted by a polynomial as
\begin{align}
   1/\xi^2
  &\sim 1.57 \times 10^{-12} + 13.55 \times \epsilon_\sigma
  - 11.16 \times \epsilon_\sigma^2
\nonumber \\
  &\propto \epsilon_\sigma~,
\label{eq:xi_fitting}
\end{align}
so $\nu=1/2$ within numerical errors.

\begin{figure}
  \includegraphics[width=8.0truecm]{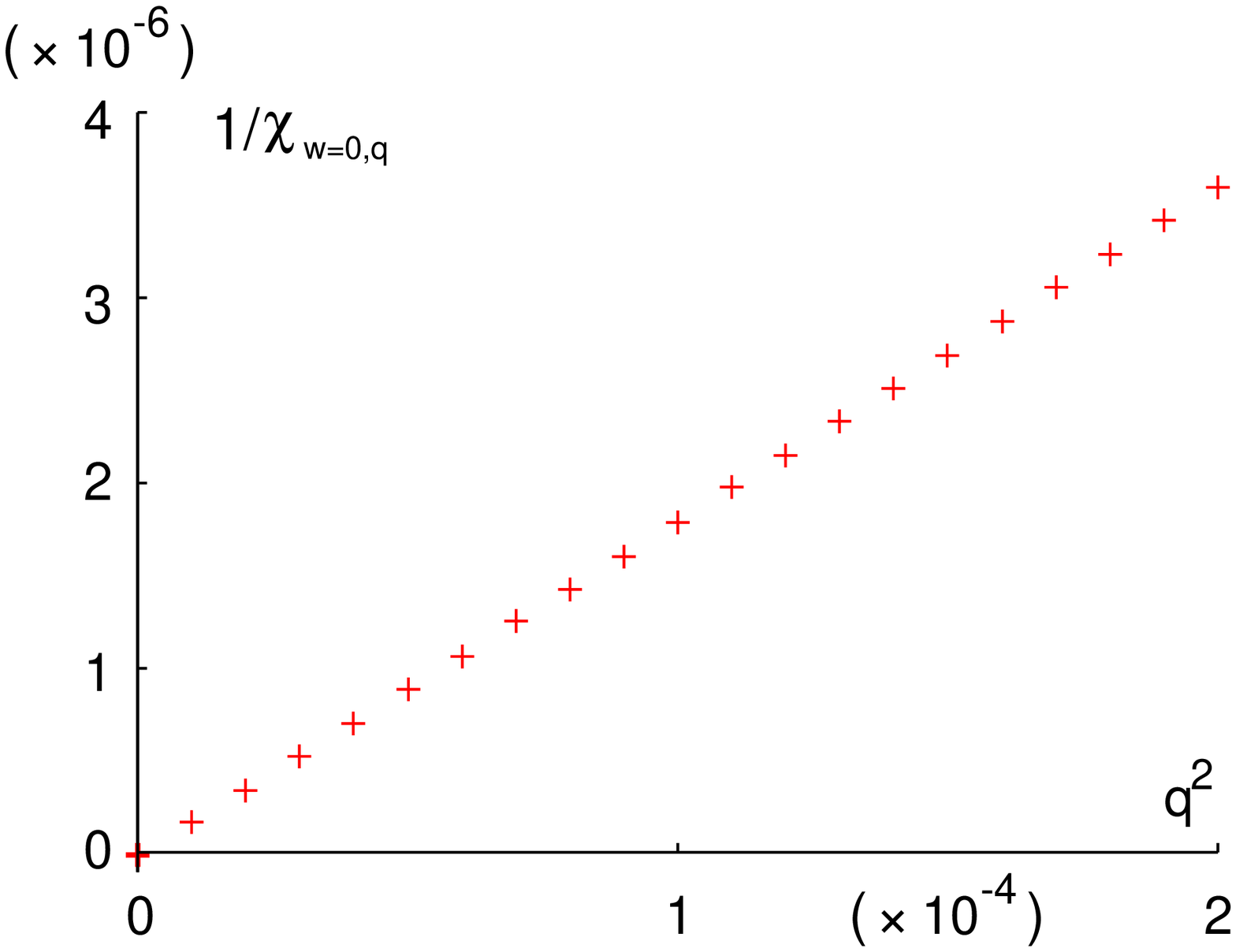}
\caption{\label{fig:eta_q} (color online).
  The static susceptibility at the critical point
  $\chi_{\mfw=0,\mfq} \vert_{T_c}$.
  Plotted are $1/\chi_{\mfw=0,\mfq} \vert_{T_c}$
  for $\mfq^2 = 10^{-5} n$ ($n = 0, 1, \cdots, 20$).
}
\end{figure}

The exponent $\eta$ comes from the static susceptibility at the critical point 
$\chi_{\mfw=0, \mfq} \vert_{T_c} \propto \mfq^{\eta - 2}$,
so it is obtained from its $\mfq$-dependence. We solve 
\begin{itemize}
\item EOM: Eq.~(\ref{eq:perturb_eq-general-varphi}) with $\mfw = 0$, 
\item Boundary condition: \eq{behavior-near_horizon-general} at the horizon,
\end{itemize}
and obtain $\chi_{\mfw=0, \mfq} \vert_{T_c}
\propto \varphi_{\mfw=0,\mfq}^{(+)}/\varphi_{\mfw=0,\mfq}^{(-)}$
from the behavior at the AdS boundary. 
Figure~\ref{fig:eta_q} shows $\chi_{\mfw=0,\mfq} \vert_{T_c}$ as a function of $\mfq$.
The result may be fitted by a polynomial as
\begin{align}
   1/\chi_{\mfw=0, \mfq} \Big\vert_{T_c}
  &\sim - 1.45 \times 10^{-8}
  + 0.018 \times \mfq^2 + 0.55 \times \mfq^4
\nonumber \\
  &\propto \mfq^2~.
\label{eq:eta_fitting}
\end{align}
so $\chi_{\mfw=0,\mfq} \vert_{T_c} \propto \mfq^{-2}$ within numerical errors.

\subsection{Relaxation time}

Finally we check $z = 2$.
This is obtained from the lowest quasinormal (QN) frequency, which gives the relaxation time. A QN frequency is obtained as an eigenvalue by solving 
\begin{itemize}
\item EOM: Eq.~(\ref{eq:perturb_eq-general-varphi}), 
\item Boundary conditions: \eq{behavior-near_horizon-general} at the horizon and $\varphi_{\mfw,\mfq}^{(-)} = 0$ (corresponding to no deformation) at the AdS boundary.
\end{itemize}
We expect that holographic superconductors belong to model A, so we set $\mfq=0$. For the other classes, one in general needs to compute $\tau_\mfq$ in order to obtain $z$. [As an example, see \eq{tau_model_B} for model B.]
\begin{figure}
  \includegraphics[width=8.0truecm]{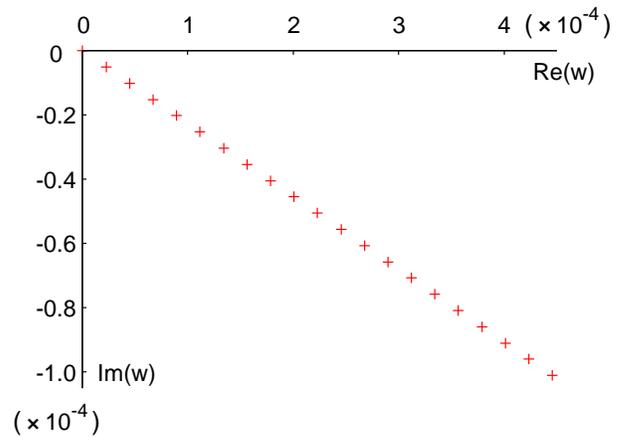}
\caption{\label{fig:QNM_q0} (color online).
  The trajectory of the lowest quasinormal frequency $\mfw_{\text{QNM}}$
  for $\epsilon_\sigma = 10^{-5} n$
  ($n = 0, 1, \cdots, 20$).
  As $\sigma \to \sigma_c$,
  $\mfw_{\text{QNM}}$ approaches the origin with equal spacing,
  and it vanishes at $\sigma = \sigma_c$.
}
\end{figure}%
Figure~\ref{fig:QNM_q0} shows the lowest QN frequency $\mfw_{\text{QNM}}$ on the complex $\mfw$-plane with the interval $\Delta\epsilon_\sigma = 10^{-5}$. 
As $\sigma$ approaches the critical value $\sigma_c$, the lowest QN frequency approaches the origin with equal spacing, and it vanishes at $\sigma = \sigma_c$. 
The equal spacing  suggests
$\mfw_{\text{QNM}} \propto \epsilon_\sigma$, 
and in fact the result may be fitted by a polynomial as
\begin{align}
   \mfw_{\text{QNM}}
  &\sim (0.76 - 1.61 i ) \times 10^{-13}
  + ( 2.23 - 0.50 i ) \times \epsilon_\sigma
\nonumber \\
  &- ( 0.52 + 0.76 i ) \times \epsilon_\sigma^2
\nonumber \\
  &\propto \epsilon_\sigma~.
\label{eq:w-q_0-fitting}
\end{align}
The imaginary part of $\mfw_{\text{QNM}}$ is nonpositive, which indicates that the system is stable in the high-temperature phase
and is marginally stable at the critical point.
Combining $\xi \propto \epsilon_\sigma^{-1/2}$
with the numerical result~(\ref{eq:w-q_0-fitting}), one obtains 
$\mfw_{\text{QNM}} \propto \xi^{-2}$ or $\tau \propto \xi^2$, namely $z = 2$.

\section{Discussion}

We find that the critical exponents for holographic superconductors take the standard mean-field values, but it is not {\it a priori} obvious. In the AdS/CFT duality, the simplest black holes with a second-order phase transition are the R-charged black holes. The $p=3$ case is dual to the ${\cal N}=4$ super-Yang-Mills theory at a finite chemical potential, and it has been widely discussed in the literature.%
\footnote{
For example, it has been used to establish the universality of the shear viscosity \cite{Mas:2006dy,Son:2006em,Saremi:2006ep,Maeda:2006by}. }
When the black holes have only one R-charge, the critical exponents take nonstandard values \cite{Cai:1998ji,Cvetic:1999rb,ref:DCP_our_1st}:
\be
(\alpha, \beta, \gamma, \delta)_{\rm R-charged} = \left( \frac{1}{2}, \frac{1}{2}, \frac{1}{2}, 2 \right)~.
\ee
While this suggests the existence of field theory systems with unconventional critical behaviors, their physical interpretation may not be easy to understand.

Then, the question is why holographic superconductors follow the conventional mean-field behavior, or how one can find a gravity dual with the conventional mean-field behavior. We do not have a completely satisfying answer, but the following argument may suggest a clue. 

Although there are six static critical exponents, only two are independent in normal statistical systems due to scaling relations. One would choose $\alpha$ as an independent exponent. The heat capacity in many interesting statistical systems does not have a power-law divergence or has only a weak power-law divergence, so $\alpha=0$. 
On the other hand, the R-charged black holes have a power-law divergence in the heat capacity, and this may be the reason why the critical exponents for these black holes do not take the conventional values. Then,  in order to have the conventional mean-field behavior, one should prepare a gravity system where the gravity sector does not have a singular behavior but only a matter sector has a singular behavior. A simple way is to pick up a familiar black hole solution without a second-order phase transition and to couple a matter field just like holographic superconductors. If the matter field undergoes a second-order phase transition, the whole system is likely to have the conventional mean-field behavior. It is clear that the gravity sector does not play an essential role in our high-temperature analysis or in the probe approximation: it simply provides a background. 

From the gravity point of view, one needs a black hole which violates the uniqueness theorem (or the no-hair theorem) in order to have a second-order phase transition. The uniqueness theorem often fails for asymptotically AdS black holes, and a simple way to violate it is again to add an appropriate matter field. 

The discussion does not ensure that any system which satisfies the above conditions has the conventional mean-field behavior because we fix only $\alpha$ among two independent critical exponents. But this discussion suggests that there may be many more gravity systems with the conventional mean-field behavior waiting to be discovered. 


\begin{acknowledgments}
We would like to thank J\"{u}rgen Berges, Misha Stephanov, and Hirofumi Wada for useful discussions. 
MN would also like to thank ``Non-equilibrium quantum field theories and dynamic critical phenomena" at the Yukawa Institute of Theoretical Physics (March 2009) where this paper is partly written. This research was supported in part by the Grant-in-Aid for Scientific Research (20540285) from the Ministry of Education, Culture, Sports, Science and Technology, Japan.
\end{acknowledgments}
%

\vspace*{0.3cm}
{\bf Note added}:

While this paper was in preparation, we received a preprint \cite{Amado:2009ts} which also studies the QNM of a holographic superconductor.)

\appendix

\section{Review of critical phenomena}\label{sec:critical}

In this appendix, we briefly review the static and dynamic critical phenomena. For more details, see standard textbooks \cite{textbook}. 


As an example, consider the ferromagnetic phase transition. In this case, the magnetization $m$ and the external magnetic field $h$ are the order parameter and a control parameter, respectively. The static critical exponents $(\alpha, \beta, \gamma, \delta, \nu, \eta)$ are defined by
\be
\begin{array}{rcll}
C_H &\propto& |\eT|^{-\alpha}~, & \\
m &\propto& |\eT|^{\beta} & (T<T_c)~, \\
\chi_T &\propto& |\eT|^{-\gamma}~, & \\
m & \propto& |h|^{1/\delta} & (T=T_c)~, \\
G(r)  & \propto& e^{-r/\xi} & (T \neq T_c)~, \\
        & \propto& r^{-d_s+2-\eta} & (T=T_c)~, \\
\xi  &\propto& |\eT|^{-\nu}~, &
\end{array}
\ee
where $\eT:= (T-T_c)/T_c$, $C_H$ is the specific heat, $\chi_T$ is the magnetic susceptibility, $G(\vec{r})$ is the correlation function, and $\xi$ is the correlation length, and $d_s$ denotes the number of {\it spatial} dimensions.

There are six static critical exponents, but not all are independent, and they satisfy static scaling relations:
\begin{subequations}
\bea
\alpha + 2\beta+\gamma =2~, 
\label{eq:scaling1} \\
\gamma = \beta(\delta-1)~, 
\label{eq:scaling2} \\
\gamma = \nu (2-\eta)~, 
\label{eq:scaling3} \\
2-\alpha = \nu d_s~.
\label{eq:scaling4} 
\eea
\end{subequations}
Because of these relations, only two are independent among six exponents, which suggests that there is some structure behind these relations, which is known as the scaling law. 

The dynamic universality class depends on additional properties of the system which do not affect the static universality class. In particular, conservation laws play an important role to determine dynamic critical exponents. As a consequence, even if two systems belong to the same static universality class, they may not belong to the same dynamic universality class. 

Phenomenologically, the relaxation to the equilibrium is governed by the time-dependent Ginzburg-Landau (TDGL) equation. A prototypical example of nonequilibrium phenomena is the Brownian motion, which is described by the Langevin equation:
\bea
\frac{dv(t)}{dt} &=& - \Gamma v(t) + \zeta(t) \\
&=& - \Gamma \frac{\del H}{\del v} + \zeta(t)~,
\eea
where $\Gamma$ is the friction coefficient, $H = v^2/2$ is the Hamiltonian, and $\zeta(t)$ is a random force with $\langle \zeta(t) \rangle =0$. The TDGL equation is the many-body generalization of the Langevin equation:
\be
\frac{\del m(t,\vecx)}{\del t} = - \int d\vecy\, \Gamma\big( |\vecx-\vecy| \big)\, \frac{\delta I[m]}{\delta m(t, \vecy)} + \zeta(t, \vecx)~,
\label{eq:TDGL}
\ee
where $\Gamma\big( |\vecx-\vecy| \big)$ is a transport coefficient which plays the role of $\Gamma$, and $I[m]$ is the pseudofree energy.

As an example, consider the Gaussian model with a source term:
\be
I[m; T, h] = \int d\vecx \,\left[ \frac{c}{2}\,(\nabla m)^2+\frac{a}{2}\,m^2 - m h \right]~,
\label{eq:gaussian}
\ee 
where $a=a_0(T-T_c)+\cdots$ and $c=c_0+\cdots$ (the dots represent the terms which do not contribute near the critical point). The static critical exponents for the Gaussian model are 
\be
(\alpha, \beta, \gamma, \delta, \nu, \eta) = \left(0, \frac{1}{2}, 1, 3,  \frac{1}{2}, 0\right)~.
\label{eq:static_exponent_LG}
\ee 
Fourier transforming \eq{TDGL} gives
\be
-i \omega \langle m_{\omega,\vecq} \rangle
= - (cq^2+a)\Gamma_q \langle m_{\omega, \vecq} \rangle 
+ \Gamma_q h_{\omega, \vecq}~.
\label{eq:TDGLq}
\ee
Here, $m_{\omega, \vecq}, h_{\omega, \vecq}, \Gamma_q$ are the Fourier components of $ m(t,\vecx),h(t, \vecx), \Gamma(|\vecx|)$, respectively. 

An interesting quantity in the critical phenomena is the response function $\chi_{\omega, \vecq}$:
\be
\chi_{\omega, \vecq} 
= \frac{ \del \langle m_{\omega, \vecq} \rangle }{ \del h_{\omega, \vecq} }~.
\ee
From \eq{TDGLq}, the response function is given by
\be
\chi_{\omega, \vecq} = \frac{ \Gamma_q }{ -i\omega+(cq^2+a)\Gamma_q }~.
\ee
The response function has a pole at $\omega=-i(cq^2+a)\Gamma_q$, which implies that the relaxation time $\tau_q$ behaves as 
\be
\tau_q^{-1} = (cq^2+a)\Gamma_q~.
\label{eq:relaxation_formula}
\ee 
In general, $\tau_{q=0}$ diverges near the critical point, which is the critical slowing down.  This divergence is parametrized by a dynamic critical exponent $z$:
\be
\tau_{q=0} \propto \xi^z~.
\ee

The dynamic universality classes are partly classified by the exponent $z$. We discuss two universality classes, model A and B. 

{\it (i) Model A:} 
In this case, $\Gamma_{q=0}$ is a nonvanishing constant. Then, \eq{relaxation_formula} gives $\tau_{q=0} \propto (T-T_c)^{-1}$. The Gaussian model has $\nu=1/2$, so $z=2$. 

{\it (ii) Model B:}
For model A, we have assumed that $\Gamma_{q=0} \neq 0$, but this is not true for a system with a conserved charge, and the value of the exponent $z$ changes as the consequence. If $\langle m_{\vecq=0}(t) \rangle$ is a conserved charge, 
\be
\frac{\del \langle m_{\vecq=0}(t) \rangle}{\del t} = 0~.
\ee
In order for this equation to be consistent with \eq{TDGLq}, $\Gamma_{q=0}=0$. The transport coefficient $\Gamma(|\vecx|)$ has even parity, so $\Gamma_q$ must be an even function of $q$. Thus, in the hydrodynamic limit, $\Gamma_q \propto q^2+O(q^4)$, and 
\be
\tau_q \propto \frac{1}{2aq^2} \propto \frac{\xi^4}{(\xi q)^2}~,
\label{eq:tau_model_B}
\ee
where the $q$-dependence is written as $(\xi q)$; this scaling form can be justified from the dynamic scaling law. Then, $z=4$.

More generally, the critical slowing down can be shown from the dynamic scaling law. 
In general, the dynamic scaling relations for model A and B are
\be
z = \left\{
	\begin{array}{ll}
	2 - \eta & \mbox{(Model A)} \\
	4 - \eta & \mbox{(Model B)}
	\end{array}
\label{eq:dynamic_static_law}
\right.
\ee
Note that \eq{dynamic_static_law} relates the dynamic critical exponent $z$ to the static critical exponent $\eta$.  For the Gaussian model, $\eta=0$, so the relations~(\ref{eq:dynamic_static_law}) reduce to the above obtained values for $z$.

We discussed only model A and B, but the dynamic universality classes were classified by Hohenberg and Halperin \cite{hohenberg_halperin}, and they are known as model A, B, C, H, F, G, and J. 
These models are further classified by the values of dynamic critical exponents.

\section{Anomalous dimension for holographic superconductors}
\label{sec:exp-crit_pt}

As we saw in the text, the critical exponents of the order parameter are given
by $\nu = \gamma/2$, $\eta = 0$, and $z = 2$,
as long as the solution of Eq.~(\ref{eq:perturb_eq-general})
can be expanded in a series of $(\mfw, \mfq)$.
Then, most of the critical exponents are determined once the critical exponent $\gamma$ is obtained.
In this Appendix, we show $\gamma=1$
by expanding the solution of Eq.~(\ref{eq:perturb_eq-general})
in a series of $\epsilon_T$.
Since $\gamma$ is a thermodynamic exponent, it is enough to consider a stationary homogeneous perturbation $\psi_{0} := \psi_{\mfw=\mfq=0}$. 

Substituting $\mfw=\mfq=0$ into Eq.~(\ref{eq:perturb-scalar_eq}),
we obtain
\begin{align}
  & 0
  = \left[ u^p \odiff{}{u} \frac{f}{u^p} \odiff{}{u}
  - l^2 m^2 \frac{H^{ \frac{2}{p-1} } }{u^2}
  \right.
\nonumber \\
  &\left. \hspace{1.5truecm}
  + \sigma^2 \frac{ H^{ \frac{2 p}{p-1} } }{f}
    \left( \frac{1}{1 + \kappa} - \frac{u^{p-1}}{H} \right)^2 \right]
  \psi_0(u)~,
\label{eq:EOM_psi_zero}
\end{align}
where we used Eq.~(\ref{eq:def-mfA}).
Also, $\sigma$ is defined in Eq.~(\ref{eq:def-sigma}) and is the control parameter for the holographic superconductors.%
\footnote{The parameter $\kappa$ is related to $\sigma$
by Eq.~(\ref{eq:sigma-kappa}).
}

At the critical point~($\sigma = \sigma_c$), 
there must exist a scalar hair solution $\bmpsi_c$ of the equation
\begin{align}
   0
  &= \left[ u^p \odiff{}{u} \frac{f_c}{u^p} \odiff{}{u}
  - l^2 m^2 \frac{H_c^{ \frac{2}{p-1} } }{u^2}
  \right.
\nonumber \\
  &\left. \hspace{1.0truecm}
  + \sigma_c^2 \frac{ H_c^{ \frac{2 p}{p-1} } }{f_c}
    \left( \frac{1}{1 + \kappa_c} - \frac{u^{p-1}}{H_c} \right)^2
   \right] \bmpsi_c(u)~,
\label{eq:EOM_bmpsi-on-crit_pt}
\end{align}
satisfying the boundary condition
\begin{align}
  & \bmpsi_c(u) \xrightarrow{u \to 0}
  \bmpsi_c^{(+)} u^{\Delta_+}~,
\label{eq:critical_sol}
\end{align}
where the subscript ``$c$" denotes the quantities at the critical point.
Note that the functional form of the electric potential $\mfA$ in the high-temperature phase 
must be valid even at the critical point by continuity.

Let us solve Eq.~(\ref{eq:EOM_psi_zero}) in a $\sigma$-expansion.
Defining 
$\epsilon_\sigma := 1 - \sigma/\sigma_c$,
$\delta\kappa := 1 - \kappa/\kappa_c$,
and $\delta\psi_0 := \psi_0 - \bmpsi_c$, we obtain
\begin{align}
   0
  &= \left[ u^p \odiff{}{u} \frac{f_c}{u^p} \odiff{}{u}
  - l^2 m^2 \frac{H_c^{ \frac{2}{p-1} } }{u^2}
  \right.
\nonumber \\
  &\left. \hspace{1.5truecm}
  + \sigma_c^2 \frac{ H_c^{ \frac{2 p}{p-1} } }{f_c}
    \left( \frac{1}{1 + \kappa_c} - \frac{u^{p-1}}{H_c} \right)^2
   \right] \delta\psi_0(u)
\nonumber \\
  & + O(\epsilon_\sigma) \bmpsi_c(u)
  + O(\delta \kappa) \bmpsi_c(u)
  + O(\epsilon_\sigma^2, \epsilon_\sigma \delta\psi_0 )~.
\label{eq:EOM_bmpsi-from-crit_pt}
\end{align}
Equation~(\ref{eq:EOM_bmpsi-from-crit_pt}) implies that 
$| \delta\psi_0 | = O(\epsilon_\sigma)$ 
since $\delta \kappa \propto \delta \sigma \propto \epsilon_\sigma$
from Eq.~(\ref{eq:sigma-kappa}). Suppose that $\delta\psi_0(u)$ behaves as
\begin{align}
  & \delta\psi_0(u)
  \sim \delta\psi_0^{(-)} u^{\Delta_-}
  + \delta\psi_0^{(+)} u^{\Delta_+}~
\label{eq:delta_psi_zero-near_boundary}
\end{align}
near the AdS boundary. Then, 
\begin{subequations}
\label{eq:psi_zero-near_boundary}
\begin{align}
  & \psi_0^{(-)} = \delta\psi_0^{(-)}
  = O(\epsilon_\sigma)~,
\label{eq:psi_zero_minus-near_boundary} \\
  & \psi_0^{(+)}
  = \bmpsi_c^{(+)} + \delta\psi_0^{(+)} \sim \bmpsi_c^{(+)}~
\label{eq:psi_zero_plus-near_boundary}
\end{align}
\end{subequations}
since $\psi_0 = \bmpsi_c + \delta\psi_0$. Combining this with Eqs.~(\ref{eq:psi_zero-eps_behavior-general}) and (\ref{eq:exponent-chi}), we obtain the critical exponent of the thermodynamic susceptibility $\gamma=1$:
\begin{align}
  & \chi
  \propto \psi_0^{(+)}/\psi_0^{(-)}
  \propto \epsilon_\sigma^{-1} \propto \epsilon_T^{-1}~.
\label{eq:TD_susceptibility-from-crit_pt}
\end{align}
%


\end{document}